\documentclass[aps,pra,twocolumn,amsmath,amssymb,showpacs,superscriptaddress,nofootinbib]{revtex4}

\newcommand{\ket}[1]{|#1\rangle}

\newtheorem{definition}{Definition}
\newtheorem{theorem}{Theorem}
\usepackage{amssymb}
\newcommand{\qed}{$\hfill \Box$}
\newcommand{\eqn}[1]{\label{#1}}
\newcommand{\eq}[1]{Eq.~(\ref{#1})}
\newcommand{\eqs}[1]{Eq.~(\ref{#1})}
\newcommand{\E}{\mathcal{E}}
\newcommand{\tr}{\mathrm{tr}}

\usepackage[dvips]{graphicx}
\usepackage{mathrsfs}

\begin{document}

\bibliographystyle{apsrev}

\title{Trade-off between the tolerance of located and unlocated errors in nondegenerate quantum error-correcting codes}
\author{Henry L. Haselgrove}
\affiliation{Defence Science and Technology Organisation,
Canberra 2600, Australia}
\affiliation{School of Information Technology and Electrical Engineering, University of New South Wales at ADFA, Canberra 2600, Australia}

\author{Peter P. Rohde}
\email[]{rohde@physics.uq.edu.au}
\affiliation{Centre for Quantum Computer Technology, Department of Physics\\ University of Queensland, Brisbane QLD 4072, Australia
}

\date{\today}

\frenchspacing

\begin{abstract}
In a recent study [Rohde et al., quant-ph/0603130 (2006)] of several quantum error correcting protocols designed for tolerance against qubit loss, it was shown that these protocols have the undesirable effect of magnifying the effects of depolarization noise. This raises the question of which general properties of quantum error-correcting codes might explain such an apparent trade-off between tolerance to located and unlocated error types.
 We extend the counting argument behind the well-known quantum Hamming bound to derive a bound on the weights of combinations of located and unlocated errors which are correctable by nondegenerate quantum codes. Numerical results show that the bound gives an excellent prediction to which combinations of unlocated and located errors can be corrected {\em with high probability} by certain large degenerate codes. The numerical results are explained partly by showing that the generalized bound, like the original, is closely connected to the information-theoretic quantity the {\em quantum coherent information}.  However, we also show that as a measure of the exact performance of quantum codes, our generalized Hamming bound is provably far from tight.
\end{abstract}


\maketitle

\section{Introduction}

Quantum computing offers the potential to solve computational problems intractable on classical computers. One of the great challenges facing the development of quantum computers is decoherence, a problem which affects all known quantum computing architectures. This has motivated much research into fault-tolerant quantum computing \cite{bib:NielsenChuang00,bib:Knill05}. The most fundamental building blocks in fault-tolerant quantum circuits are quantum error correcting codes (QECC's) \cite{bib:Shor95,bib:Steane96,bib:CalderbankShor96,bib:Calderbank97}, which encode logical qubits in a way that tolerates some number of physical errors.

In certain physical realizations of quantum computing, errors of two distinct types will be present: \emph{located} and \emph{unlocated} errors. A located error is one that leaves behind a classical signal indicating which qubit was affected. An unlocated error, on the other hand, corrupts the state of a qubit without leaving any such additional evidence. An important example of a located error is qubit loss; for example the loss of a photon in optical quantum computing\footnote{In principle, one can use a quantum non-demolition measurement to non-destructively test for the presence of a photon.}. (We use the terms {\em located error} and {\em loss} interchangeably in this paper. Loss errors are also known as {\em erasure} errors in some contexts, e.g. \cite{Grassl97a}.)


Traditionally QECC's have focused on correcting unlocated errors. Specifically, most existing codes and protocols aim to protect against unlocated depolarizing noise (where each qubit with some probability enters the maximally mixed state). Recently, however, especially with the advent of photonic quantum computing architectures \cite{bib:KLM01,bib:Kok05}, several codes and protocols have been suggested for dealing specifically with located errors in the form of qubit loss \cite{bib:RalphHayes05,bib:Varnava05}. It has recently been shown \cite{bib:RohdeRalphMunro06b} that these loss-tolerant protocols have the negative side-effect of amplifying depolarizing noise. Since scalable quantum computing requires tolerance against both error types, it is important to understand what type of fundamental constraints exist on the ability of quantum error-correcting codes to correct combinations of the two noise types.

In this paper we consider a generalization of the quantum Hamming bound. For any nondegenerate quantum error-correcting code, our generalized bound provides an upper limit on the values $(t_l, t_u)$ such that any error pattern consisting of $t_l$ located errors and $t_u$ unlocated errors can be corrected by the code. We study a number of aspects of the behaviour of this bound. We compare the bound to the performance of large concatenated codes, and show that the bound gives a very tight correspondence to the maximum values $(t_l, t_u)$ such that {\em most} patterns of $t_l$ located errors and $t_u$ unlocated errors can be corrected by the code. That is, the bound appears to predict which located and unlocated error weights are correctable with {\em high probability} by certain codes.
Despite this, we show that the bound is in fact quite loose in bounding those error weights that can be corrected with {\em certainty} by a code. We also consider the behaviour of the bound in the limit of large code size, and show that it reduces to a simple condition relating to the coherent information of the error channel, a quantity that is known to be closely connected to the quantum capacity of a noisy channel.

In the remainder of this introductory section, we briefly review nondegenerate quantum error-correcting codes and the quantum Hamming bound. Then in the section that follows, the generalized bound is introduced.


\textbf{Nondegenerate quantum error-correcting codes and the quantum Hamming bound:}
A QECC that encodes $k$ qubits into $n$ qubits is a $2^k$-dimensional subspace of the $2^n$-dimensional state space on $n$ qubits. Denote this subspace $V$. A quantum code that corrects $t$ unlocated errors is said to be nondegenerate if each of the subspaces $\sigma V$ are orthogonal to one another, where $\sigma$ takes the value of all possible $n$-qubit tensor products of Pauli operators having weight\footnote{The {\em weight} of $\sigma$ is the number of non-identity terms in the tensor product.} at most $t$. That is, the code can distinguish any correctable Pauli error from the others, since it is possible to perform a measurement to determine which of the orthogonal subspaces $\sigma V$ has been entered. Hence, by the result known as the {\em discretization of errors} (\cite{bib:NielsenChuang00}, Sec.~$10.3.1$), all errors affecting $t$ or less qubits (whether Pauli or not) will be correctable by the code.

The quantum Hamming bound \cite{Ekert96b} expresses the fact that, since the subspaces $\sigma V$ are all orthogonal to one another, the sum of the dimensions of the $\sigma V$ cannot exceed the number of dimensions of the entire state space of $n$ qubits.
There are $\sum_{i=0}^t \binom{n}{i} 3^i$ distinct $n$-qubit tensor products of Pauli operators that have weight at most $t$ . The $3^i$ term arises from the fact that at every location there are three non-identity Pauli errors that may occur -- $X$, $Y$ and $Z$. Each subspace $\sigma V$ has dimensionality $2^k$, and the total state space has dimensionality $2^n$, and hence the quantum Hamming bound is:
\begin{equation}
\sum_{i=0}^t \binom{n}{i} 3^i 2^k \leq 2^n. \eqn{ohb}
\end{equation}
Interestingly, this bound does not make any assumptions about the nature of the encoding or recovery operations. Instead it is based purely on a counting argument. While the Hamming bound does not cover degenerate quantum codes, of which there are many examples, it provides some insight into the behaviour of QECC's in general -- to date no quantum codes are known to violate the Hamming bound \cite{bib:NielsenChuang00}.

\section{The generalized quantum Hamming bound}
 We now modify the quantum Hamming bound to accommodate for error patterns which are a combination of unlocated and located errors:
 \begin{theorem}
  If a nondegenerate code encoding $k$ logical qubits into $n$ qubits corrects all error patterns consisting of at most $t_u$ unlocated and $t_l$ located errors, then the following generalized quantum Hamming bound holds:
\begin{equation} \label{eq:gen_hamm_bound}
4^{t_l}\sum_{i=0}^{t_u} \binom{n-{t_l}}{i} 3^i 2^k \leq 2^n.
\end{equation} \label{thm:ghb}
\end{theorem}
{\bf Proof:} Assume that $t_l$ located errors have occurred (and that the location and number of these errors are known to the decoder), and that in addition no more than $t_u$ unlocated errors have occurred on the remaining $n-t_l$ qubits. The code is nondegenerate, so for reliable decoding to occur the subspaces $\sigma V$ must be orthogonal to one another, where $\sigma$ ranges over all $n$-qubit Pauli tensor products that have weight at most $t_u$ on the $n-t_l$ qubits where located errors have not occurred (and arbitrary weight on the $t_l$ qubits where located errors have occurred). There are $4^{t_l} \sum_{i=0}^{t_u} \binom{n-{t_l}}{i} 3^i$ such values of $\sigma$. Each subspace $\sigma V$ has dimension $2^k$, and the total state-space has dimension $2^n$, hence \eq{eq:gen_hamm_bound} follows.\qed

\begin{figure}
\includegraphics[width=0.55\columnwidth]{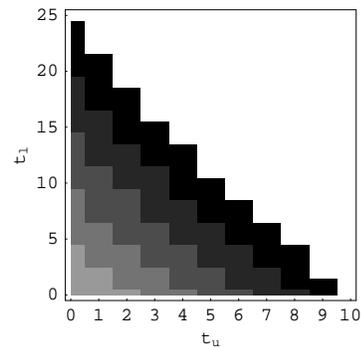}
\caption{Shows the values of $(t_u,t_l)$ which satisfy the generalized quantum Hamming bound, for $n=50$ (black), $n=40$, $n=30$, $n=20$ and $n=10$ (light gray) codes.} \label{fig:error_tradeoff}
\end{figure}

It is insightful to consider the limiting behavior of this modified bound. In the limit where no located errors have occurred, the bound simply reduces to the original quantum Hamming bound, as expected. In the opposing limit, where \emph{only} located errors occur, the inequality reduces to $t_l\leq(n-k)/2$. This bound reaffirms the well-known no-cloning limit and represents an intuitive upper bound\footnote{The no-cloning theorem states that for an arbitrary unknown state $\ket\psi$, it is impossible to perform the transformation $\ket\psi\to\ket\psi\otimes\ket\psi$, i.e. to make two identical copies of the state. To see how this relates to the number of located errors one can correct for, consider the following. Suppose we encode a single logical qubit into an $n$ qubit codeword. If we divide the codeword in two and give each half to a different party, both parties would be able to reproduce the original codeword if they could correct $n/2$ or more located errors, since they are both missing half their qubits. Thus, for $k=1$, $t_l\leq(n-1)/2$. For $k>1$, consider the following. Suppose our encoding operation maps the first $k-1$ logical qubits to the first $k-1$ codeword qubits directly, and the remaining logical qubit to the remaining $n-k+1$ codeword qubits. This strategy maximizes our ability to correct errors on the last logical qubit, assuming a unitary encoding operation. Now if $t_l\geq(n-k+1)/2$ we could clone the last logical qubit. Thus, $t_l\leq(n-k)/2$.}. The bound is shown graphically for some small values of $n$ in Fig.~\ref{fig:error_tradeoff}.

\section{large-$n$ limit}
In this section we derive the large-$n$ limiting form of the generalized Hamming bound, expressed as a function of the error rates $t_l/n$ and $t_u/(n-t_l)$.

From \eq{eq:gen_hamm_bound} it follows that
\begin{equation}
4^{t_l}  {{n-t_l}\choose{t_u}} 3^{t_u} 2^k \leq 2^n, \label{ghb2}
\end{equation}
where all terms other than $i=t_u$ in the sum have been dropped. In the large-$n$ limit the $i=t_u$ term dominates the left hand side of \eq{eq:gen_hamm_bound}, so \eqs{ghb2} and (\ref{eq:gen_hamm_bound}) are essentially equivalent in this limit.

Taking the logarithm (base 2) of both sides of \eq{ghb2}, and
substituting ${A\choose B} = \frac{A!}{B!(A-B)!}$ gives
\begin{equation}
2t_l + \log(n-t_l)! - \log t_u! - \log (n-t_l-t_u)! + t_u \log 3 + k
\leq n \eqn{fac_prev}
\end{equation}


Now, Stirling's approximation states that for large $N$,
\begin{equation}
\log N! \approx N \log N - N \log e.
\end{equation}
Using this approximation, \eq{fac_prev} becomes (after some
simplification, and after dividing both sides by $n$)
\begin{eqnarray}
2q - p(1-q)\log p + p(1-q)\log 3 \hspace{3cm} \nonumber \\   - (1-q)(1-p) \log(1-p) -1 + r \lesssim 0,
\label{eq2}
\end{eqnarray}
where $q\equiv t_l/n$, $p \equiv t_u/(n-t_l)$, and $r \equiv k/n$. The variables $q$ and $p$ can be interpreted as the rate of unlocated errors and located errors respectively. (The appropriateness of the denominator $n-t_l$ in the definition of $p$ is explained by the fact that if  both a located and unlocated error is by coincidence applied to the same qubit then the result is simply a located error). $r$ is the information rate of the code (that is, the average number of encoded qubits transmitted per physical qubit sent through the channel).

\eq{eq2} has a particularly simple form when expressed in terms of {\em coherent information} \cite{Schumacher96b}. The coherent information is a measure of how well
a channel $\E$ preserves quantum correlations that exist between a system $Q$ and other auxiliary systems when $Q$ is sent through $\E$.
\begin{definition}[Coherent Information] \label{def:ci}
Let $\rho$ be the state of some system $Q$. Let $R$ be an auxiliary
system which purifies $\rho$. That is, there is some pure joint state
$|\psi\rangle$ of $Q$ and $R$ such that $\rho = \tr_R
(|\psi\rangle\langle \psi|)$. Say that the channel $\E$ is applied to
system $Q$ only, resulting in the a joint state $(\E \otimes I)
(|\psi\rangle\langle \psi|)$ on $QR$ and the reduced state $\E(\rho)$
on $Q$. By definition, the coherent information equals
$$ I(\rho,\E) \equiv S[\E(\rho)] - S[(\E \otimes I) (|\psi\rangle\langle
\psi|)],
$$ where $S[\cdot]$ denotes the von Neumann entropy.
\end{definition}

Consider a channel $\epsilon$, parameterized by $p$ and $q$, that has the following effect on a qubit: with probability $p$ an unlocated depolarizing error is applied, and independently with probability $q$ a located depolarizing error is applied. That is, the channel modifies an arbitrary qubit state $\sigma$ as follows:
\begin{eqnarray}
\E(\sigma \otimes |NL\rangle\langle NL|) = (1-q)(1-p) \sigma \otimes |NL\rangle\langle NL| && \hspace{5mm} \nonumber \\
 \hspace{5mm} + (1-q)\frac{p}{2} I \otimes |NL\rangle\langle NL|
+ \frac{q}{2} I\otimes |L\rangle\langle L|,&&
\end{eqnarray}
where the states $|NL\rangle$ and $|L\rangle$ represent the classical signal indicating whether a located error has occurred, and where $I$ is the one-qubit identity matrix.

 Suppose that $\sigma$ is equal to one-half of a maximally-entangled qubit pair (that is,  $\sigma$ is a maximally mixed state). Then it is straightforward to evaluate the two required von Neumann entropy values in Definition~\ref{def:ci}, giving
$S[\E(\rho)]$ $=$ $H[(1-q)(1-p), \textstyle\frac{1}{3}(1-q)p, \frac{1}{3}(1-q)p,
\frac{1}{3}(1-q)p, \frac{1}{4}q, \frac{1}{4}q, \frac{1}{4}q,
\frac{1}{4}q]$ and $S[(\E\otimes I)(|\psi\rangle\langle\psi|)]$ $=$ $H[\textstyle\frac{1}{2}(1-q), \frac{1}{2}(1-q), \frac{1}{2}q,
\frac{1}{2}q]$, where $H[\cdot]$ denotes the Shannon entropy of a probability distribution. By utilizing the formula for Shannon entropy, one obtains the following expression for the coherent information of the maximally-mixed state sent through the depolarizing channel with unlocated/located error rates $(p,q)$:
\begin{eqnarray}
I(\rho,\E) = 1+(1-q)(1-p)\log(1-p) \hspace{2cm} && \nonumber \\
\hspace{2cm} \phantom{0}+ p(1-q) \log p - p(1-q)\log 3 - 2q.&&
\end{eqnarray}
Comparing with \eq{eq2}, we see that the large-$n$ limit of the generalized Hamming bound can be written succinctly as
\begin{equation}
I(\rho,\E)\ge r.
\end{equation}
Note that in \cite{Adami96a} a similar relation was found between the asymptotic form of the original quantum Hamming bound and the coherent information of a unlocated depolarizing channel. This indicates that our generalization of the quantum Hamming bound is in some sense a natural one.

\section{Numerical results: large concatenated codes}

In this section we give numerical results which show that there exist codes whose performance against combinations of unlocated and located noise is closely governed by the generalized quantum Hamming bound. The codes we consider are created by concatenating the 5 qubit code with itself several times. These codes are degenerate, and so {\em a priori} do not necessarily satisfy a Hamming bound. However these codes have decoding algorithms that are optimal and efficient \cite{Poulin06a}, making them amenable to numerical study.

Poulin's method \cite{Poulin06a} for efficiently performing the maximum-likelihood decoding and correction of a $5^L$-qubit multiply-concatenated code can be described briefly as follows. Input to the decoder are the {\em a priori} error distributions of each qubit. That is, for each $i=1,\dots,5^L$ we have a vector $[p^{(i)}_I,p^{(i)}_X,p^{(i)}_Y,p^{(i)}_Z]$ which represents the probability of each of the four Pauli errors affecting qubit $i$. For example, if the $i$-th qubit is known to have experienced a located depolarization error, the $i$-th input distribution will equal $[0.25,0.25,0.25,0.25]$, otherwise it will equal $[1-p,p/3,p/3,p/3]$, where $p$ is the rate of unlocated depolarization errors. The decoder consists of $L$ ``passes''. In the first pass, the decoder corrects each of the $5^{L-1}$ 5-qubit blocks separately. For each 5-qubit block, this step consists of measuring the syndrome $s$ of that block with respect to the 5-qubit code, applying a recovery operation $R$ which consists of some pattern of Pauli operators whose syndrome matches $s$,\footnote{For any given syndrome $s$ there are 64 such patterns; it doesn't matter which one is chosen.} and outputting the posterior error probability distribution $[p_I,p_X,p_Y,p_Z]$. The posterior error probability distribution describes the probability that the combined effect of the error pattern and recovery pattern on a given block corresponds to either an encoded $I$, $X$, $Y$, or $Z$ operation on that block. (The values are straightforward to calculate, given knowledge of the prior distributions $[p^{(i)}_I,p^{(i)}_X,p^{(i)}_Y,p^{(i)}_Z]$, the recovery $R$, and the stabilizers and logical operators of the code.) Thus, for each of the $5^{L-1}$ code blocks we have a distribution over encoded errors. In the second pass, these are reinterpreted as the {\em a priori} error distributions on a series of $5^{L-1}$ physical qubits, and the entire procedure above is repeated, giving the distribution of errors on a series of $5^{L-2}$ blocks at the next level of decoding.
When the procedure has been carried out for $L$ passes in total, we are left with a distribution over the four possible logical errors of the single $5^L$-qubit code block. The logical operation with the highest probability is selected, and applied to the state as a recovery operation. This completes the maximum-likelihood decoding and correction of the input state.


\begin{figure}
\includegraphics[width=.9\columnwidth]{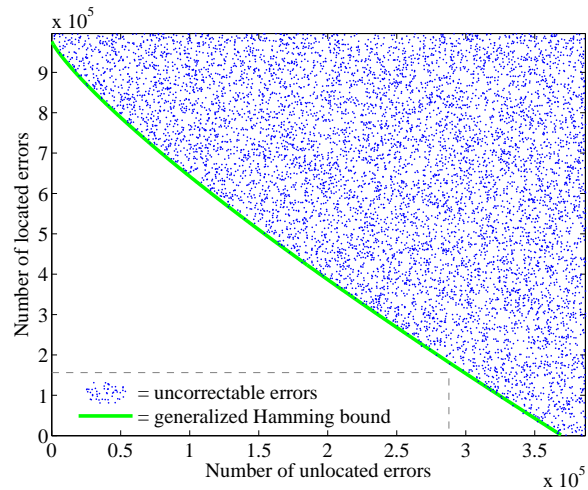}
\caption{A plot of those randomly-chosen error weights which caused a decoding failure for the $5^9$-qubit code. Also shown for comparison is the location of the generalized quantum Hamming bound. }\label{fig:scatter}
\end{figure}
Fig.~\ref{fig:scatter} shows the results of simulating the above procedure for a range of different located and unlocated error weights, for the $5^9=1953125$-qubit code. The simulation consisted of approximately 20000 trials. For each trial, the number of unlocated errors ($t_u$) and the number of located errors ($t_l$) were chosen randomly from the entire plot area. A random Pauli error pattern was then chosen, consistent with the chosen values $t_u$ and $t_l$. Maximum-likelihood syndrome decoding was then performed, and a point was plotted on the figure whenever the decoding failed. The results show that the weights of the uncorrectable errors form a region that very closely resembles the generalized quantum Hamming bound. Similar results (not shown) were achieved with a multiply-concatenated $7$ qubit code.

However, the results do not imply that the bound is ``tight'' in the usual sense. This is because the results of Fig.~\ref{fig:scatter} indicate a different property of the code to that which the quantum Hamming bound indicates. The numerical results show which error weights have some reasonable probability of being uncorrectable. On the other hand, the quantum Hamming bound places a limit on which error weights are correctable with certainty. To illustrate the point, note that there is a simple argument to show that there exists a Pauli error pattern having $(t_u,t_l)=(9842,0)$ that is uncorrectable by the $5^9$-qubit code. This point is far inside the Hamming bound shown in Fig.~\ref{fig:scatter}. However, a randomly chosen $(t_u, t_l)=(9842,0)$ error pattern will have an extremely small probability of being uncorrectable, hence it will not be encountered in experiments such as Fig.~\ref{fig:scatter}.

\section{Tighter bound}

In \cite{Grassl97a}, the following simple general property of QECCs was observed: a QECC can correct all weight-$t$ located errors if and only if it can correct all weight-$2t$ unlocated errors. This result can be generalized to the following theorem, which in turn provides a way to tighten the generalized quantum Hamming bound.

\begin{theorem}
A quantum error-correcting code can correct all patterns of $t$ unlocated errors if and only if it can correct all patterns of errors that are a combination of $2m$ located errors and $t-m$ unlocated errors. This statement holds for all integers $0\le m \le t$. \label{thmone}
\end{theorem}
{\bf Proof:}
Let $L$ be a set of $2m$ locations within the code. Let $\{E^{(L,t-m)}_i\}$ be the set of all Pauli operators that act as the non-identity on at most $t-m$ qubits outside the set $L$ (but with no restriction on how they act on the qubits in the set $L$).
From the quantum error-correcting conditions (\cite{bib:NielsenChuang00}, Sec.~$10.3$) it follows that the code $C$ can correct all combinations of $2m$ located errors and $t-m$ unlocated errors if an only if
\begin{equation}
PE^{(L,t-m)\dagger}_iE^{(L,t-m)}_jP = \alpha^{(L,m)}_{ij} P \hspace{3em}\mbox{(for all $i,j,L$)},
\end{equation}
for some complex numbers $\alpha^{(L,m)}_{ij}$, and where $P$ projects onto the codespace.

Now, clearly the set of products $\{E^{(L,t-m)\dagger}_iE^{(L,t-m)}_j\}$ just corresponds to the set $\{E^{(L,2(t-m))}_i\}$, that is the set of Pauli operators which act as the non-identity on at most $2(t-m)$ qubits outside the set $L$ (but with no restriction on how they act on the qubits in the set $L$). So, the above condition can be written equivalently as
\begin{equation}
PE^{(L,2(t-m))}_iP = \beta^{(L,m)}_{i} P \hspace{3em}\mbox{(for all $i,L$)}, \eqn{eq:pep}
\end{equation}
for some complex $\beta^{(L,m)}_i$.

Now, each $E^{(L,2(t-m))}_i$ acts as the non-identity on at most $2m+2(t-m) = 2t$ qubits. In fact, the set $\{E^{(L,2(t-m))}_i\}$ over all possible $i$ and $L$ is the entire set of Pauli operators that act on at most $2t$ qubits, which we denote $\{E^{(\emptyset,2t)}_i\}$. Then \eq{eq:pep} can be written equivalently as
 \begin{equation}
PE^{(\emptyset,2t)}_iP = \gamma_{i} P \hspace{3em}\mbox{(for all $i$)}, \label{eq:cond}
\end{equation}
for some complex $\gamma_i$.
Thus, the statement ``a code can correct all patterns of errors that are a combination of $2m$ located depolarization errors and $t-m$ unlocated depolarization errors'' holds if and only if the condition in Eq.~(\ref{eq:cond}) holds. But Eq.~(\ref{eq:cond}) does not depend on $m$. Thus the different versions of the statement for the various values of $m$ are all equivalent to one other (since they are each equivalent to \eq{eq:cond}), including the one where $m$ is set to zero. This completes the proof. \qed

\begin{figure}
\includegraphics[width=\columnwidth]{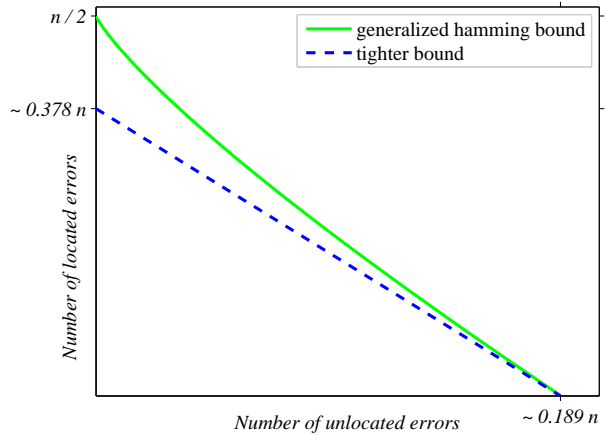}
\caption{An illustration of the generalized quantum Hamming bound, in comparison to the tighter bound that is obtained using Theorem~\ref{thmone}, for large rateless codes. } \label{fig:tight}
\end{figure}

Fig.~\ref{fig:tight} illustrates how Theorem~\ref{thmone} can be used to generate a bound which is significantly tighter than the generalized quantum Hamming bound, in the case of large codes having an asymptotically zero rate (that is, $k/n \approx 0$, such is the case for large multiply-concatenated codes). No nondegenerate QECC can exist which corrects all errors of a weight corresponding to a point above the dashed line. If such a code did exist, Theorem~\ref{thmone} would imply that the same code would break the original quantum Hamming bound (that is, be able to correct a number of unlocated errors corresponding to a point on the $X$-axis of Fig.~\ref{fig:tight} to the right of where the solid line intersects), thus giving a contradiction.
The tighter bound can be stated formally as follows:
\begin{theorem}
  If a nondegenerate code encoding $k$ logical qubits into $n$ qubits corrects all error patterns consisting of at most $t_u$ unlocated and $t_l$ located errors, then the following bound holds:
\begin{equation} \label{eq:tighter_bound}
\sum_{i=0}^{t_u+\lfloor t_l/2\rfloor } \binom{n}{i} 3^i 2^k \leq 2^n
\end{equation} \label{thm:tighter}
\end{theorem}
{\bf Proof:} By Theorem~\ref{thmone}, the code can correct all patterns of $t_u+\lfloor t_l/2\rfloor$ unlocated errors. Thus, the quantum Hamming bound (\eq{ohb}) applies, with $t=t_u+\lfloor t_l/2\rfloor$, hence \eq{eq:tighter_bound}. \qed

Thus, although it would appear that dimension-counting arguments in the form of the generalized quantum Hamming bound can give an accurate indication of the located and unlocated error weights which can be corrected with {\em high probability} by certain codes, in general it provides a poor indication of which error weights can be corrected exactly.

\section{Discussion \& conclusion}

We have derived two versions of a bound on the number of unlocated and located errors correctable by nondegenerate QECCs. The first was derived using dimension-counting arguments in a direct generalization of the quantum Hamming bound. It was seen to be an accurate prediction of the performance of large concatenated codes against random combinations of located and unlocated errors. This is likely to extend to most typical large codes, due to the asymptotic connection between the bound and the ``coherent information'', which in turn is closely related to the performance of ``random hashing'' error-correcting protocols.

A significantly tighter version of the bound was derived by applying the quantum error-correction conditions. Thus, it would appear to be a general feature of QECCs that the set of unlocated and located error weights correctable with certainty differs significantly to the set correctable with high probability.

Both forms of the bound show a general ``trade-off'' between a code's ability to simultaneously correct both unlocated and located errors. However, the trade-off is well-behaved: requiring a code to correct a small number of unlocated errors will only have a small impact of the code's ability to simultaneously correct located errors.

Thus, the high sensitivity of the various loss-tolerant protocols to unlocated noise is not likely to be attributable to general properties of QECCs, but rather to particular features of the protocols employed. Rather, we speculate that as a consequence of Theorem~\ref{thmone} the loss-specific codes such as parity codes \cite{bib:RalphHayes05} and horticultural codes \cite{bib:Varnava05} will also have high tolerance to unlocated noise, when considered apart from the protocols in whose context they were defined.

\begin{acknowledgments}
We thank Tim Ralph, Bill Munro, Michael Nielsen, Alastair Kay, and Chris Dawson for helpful discussions and feedback. PPR acknowledges support by the Australian Research Council, Queensland State Government, and the DTO-funded U.S. Army Research Office Contract No. W911NF-05-0397.
\end{acknowledgments}


\end{document}